\documentclass{bioinfo}
\copyrightyear{2013}
\pubyear{2013}

\usepackage{url}
\usepackage{graphicx}
\usepackage{multirow}
\usepackage{threeparttable}
\usepackage[multiple]{footmisc}

\begin{document}
\firstpage{1}

\title[Adaptive quality score compression]{Adaptive reference-free compression of sequence quality scores}
\author[Janin \textit{et~al.}]{Lilian Janin,$^{1}$ Giovanna Rosone\,$^{2}$ and Anthony J. Cox$^{1,}$\footnote{to whom correspondence should be addressed}}
\address{$^{1}$Computational Biology Group, Illumina Cambridge Ltd., Chesterford Research Park, Little Chesterford, Essex CB10 1XL, United Kingdom\\
$^{2}$University of Palermo, Dipartimento di Matematica e Informatica,Via Archirafi 34, 90123 Palermo, Italy\\}

\history{Received on XXXXX; revised on XXXXX; accepted on XXXXX}

\editor{Associate Editor: XXXXXXX}

\maketitle

\begin{abstract}

\section{Motivation:}

Rapid technological progress in DNA sequencing has stimulated interest in compressing the vast datasets that are now routinely produced. Relatively little attention has been paid to compressing the quality scores that are assigned to each sequence, even though these scores may be harder to compress than the sequences themselves.
By aggregating a set of reads into a compressed index, we find that the majority of bases can be predicted from the sequence of bases that are adjacent to them and hence are likely to be less informative for variant calling or other applications. The quality scores for such bases are aggressively compressed, leaving a relatively small number at full resolution. Since our approach relies directly on redundancy present in the reads, it does not need a reference sequence and is therefore applicable to data from metagenomics and de novo experiments as well as to resequencing data.

\section{Results:}

We show that a conservative smoothing strategy affecting 75\% of the quality scores above Q2 leads to an overall quality score compression of 1 bit per value with a negligible effect on variant calling.
A compression of 0.68 bit per quality value is achieved using a more aggressive smoothing strategy, again with a very small effect on variant calling.

\section{Availability:}

Code to construct the BWT and LCP-array on large genomic data sets is part of the \texttt{BEETL} library, available as a github respository at \url{git@github.com:BEETL/BEETL.git}.

\section{Contact:} \href{acox@illumina.com}{acox@illumina.com}

\end{abstract}

\def\bwtS#1{\textsf{bwt}_{#1}(\mathcal{S})}
\def\lcpS#1{\textsf{lcp}_{#1}(\mathcal{S})} %
\def\bwtw#1{\textsf{bwt}_{#1}({w})}
\def\lcpw#1{\textsf{LCP}_{#1}({w})}
\def\bigS{\mathcal{S}}
\def\BCR{\texttt{BCR} }
\def\BWT{\textsf{BWT}} %
\def\LCP{\textsf{LCP}} %
\def\SA{{\textsf SA}} %
\def\BCRLCP{\texttt{extLCP} }
\def\BWTE{\texttt{bwte} }
\def\sort#1{\textrm{sort}(#1)}
\def\BEETL{\texttt{BEETL}}
\def\bzip2{\texttt{bzip2}}

\def\gzip{\texttt{gzip}}
\def\7zip{\texttt{7-Zip}}
\def\BCRext{\texttt{BCRext} }
\def\recoil{\texttt{ReCoil}}

\section{Introduction}\label{sec:intro}

The raw output of a DNA sequencer is converted by a program known as a \emph{base caller} into nucleotide bases, each of which is typically assigned a \emph{quality score} that estimates the probability that the base has been sequenced correctly. Quality scores have long been used to trim the low quality ends of reads and for accurate consensus sequence determination
(\cite{BonfieldStaden1995,Marth1999} ).
More recently they have enabled more accurate alignments of the shorter sequences produced by ``next-generation'' technologies, by allowing the aligner to give lower weight to mismatches at less reliable base positions (\cite{Smith2008,LiJueDurbin2008}).
Often quality scores are expressed on an integer scale derived from the error probability $p$ via the formula $-10\textrm{log}_{10}p$, a scoring scheme named after the Phred base caller (\cite{EwingGreen1998}) that first employed it.

The widely-used FASTQ format (\cite{Cock2010}) stores the sequence and metadata of a set of DNA reads as ASCII text, together with one-character-per-score \emph{quality strings} that encode the Phred scores of their bases. The different properties of these three data types have meant that many FASTQ compression methods have treated them as three distinct data streams and applied separate compression strategies to each.

The metadata field tends to be formatted in ways that are specific to the technology that was used to generate the sequence and some FASTQ compressors have exploited such structure to improve compression. However, there is no global format specification for the metadata so any universally applicable method for its compression must necessarily be a generic exercise in the compression of ASCII text.

While standard text compressors such as \gzip\footnote{\texttt{www.gzip.org}, Jean-loup Gailly and Mark Adler} do not significantly outperform a naive 2-bits-per-base encoding on DNA sequence data, applications such as resequencing and de novo assembly typically rely on a 20-fold or more oversampling of the underlying genome and this redundancy can be exploited to improve compression of the sequences themselves. \emph{Reference-based compression} tools such as CRAM (\cite{Fritz2011}) encode reads in terms of differences between their sequences and the sites they align to on a reference sequence. Sorting the reads by the coordinates of these alignments saves most of the overhead of storing their positions and is a convenient ordering for applications such as SNP calling and visualization.

Despite these advantages, reference-based compression suffers when the reference is incomplete (reads that do not align cannot be compressed), subject to change or not present at all (as in metagenomics), motivating an interest in \emph{reference-free} compression methods. The tool QUIP (\cite{Jones2012}) creates an on-the-fly de novo assembly to perform reference-based compression against, while SCALCE (\cite{HachNumanagicAlkanSahinalp2012}) places similar reads near to each other in a sorted file, facilitating good performance by standard tools such as \texttt{gzip} that operate on a buffer of text at a time.

Another widely-used generic compression tool \bzip2\footnote{\texttt{www.bzip.org}, Julian Seward} exemplifies \emph{Burrows-Wheeler transform (BWT) compression}: text is split into 900Kb blocks and the BWT of each block is computed. The BWT is a reversible permutation of the text that acts as a \emph{compression booster} for the pipeline of standard compression steps that \bzip2 subsequently applies. In \cite{CoxBauerJakobiRosone2012}, two of the present authors showed that, while \bzip2 performs comparably to \gzip\ on DNA sequence reads, the compression achieved by BWT-based methods improves by more than threefold if the BWT of the entire read set is built, since this captures redundancy between reads that were widely spaced in the original file. It was shown that compression can be further boosted by pre-sorting the reads or applying an implicit sorting strategy while the BWT is being built, enabling compression of better than 0.5 bits-per-base to be achieved.

Lossless approaches to the compression of quality scores have exploited empirical relationships between the scores assigned to bases within a read: for instance, Illumina quality scores tend to be monotone decreasing along a read with a decrease in scores between adjacent bases that is usually small. An over-reliance on such observations potentially ties a compression scheme to a given sequencing technology and makes it sensitive to changes in sequencing protocol.
Moreover, it is likely that Illumina quality scores at their full resolution contain a proportion of random noise that is impossible to compress: striking evidence for this is given by Table 4 in \cite{BonfieldMahoney2013}, which shows multiple entrants to the SequenceSqueeze competition for FASTQ compression achieving similar lossless compressions of around 2.94 bits-per-score on a test dataset, but that no entrants were able to improve upon this figure.

It is undesirable that, when compressed, the quality scores should take up several times more space than the sequences themselves and so we are led to consider compressing them in a \emph{lossy} way. \cite{Kozanitis2010} found that a \emph{global} reduction in the resolution of the scores from 40 values  to 8 values (thus permitting each score to be stored in 3 bits) had no significant impact on the quality of variant calls, while strategies for global requantization of quality scores were studied in more detail by \cite{Wan2012}.

However, treating all scores in the same way ignores the fact that most of them could likely be reduced in resolution or even discarded entirely with little impact upon our ultimate goal of ensuring that analyses performed with the reduced scores closely reflect the results obtained from the original data. In a human resequencing context, for example, if a large coverage of high-quality bases unanimously supports a homozygous match to the reference genome then a confident call can be made without the full-resolution quality scores of each individual base needing to be kept. With this in mind, CRAM allows an \emph{adaptive} approach where only quality scores that contribute to variant calls that do not match the reference are kept. This enables the vast majority of scores to be omitted but means compression cannot take place until analysis has been finalized. This means any pre-analysis transfer or storage of the data will not benefit from compression and is potentially problematic if the data subsequently needs to be reanalysed.

Here we present an adaptive and reference-free approach to lossy quality-score compression. Our central premise is that if a base in a read can, with high probability, be \emph{predicted} by the \emph{context} of bases that are next to it, then the base itself is imparting little additional information and its quality score can be discarded or aggressively compressed at little detriment to downstream analysis. Such predictions are made by considering all possible contexts present in the reads: if every occurrence of some string $Q$ is followed by the same character $c$ then the presence of a context $Q$ in a read can be said to predict that $c$ will come next.

In the rest of this paper we formalize this intuition and give algorithms that use the BWT of a set of reads to identify non-essential quality scores. The BWT places all characters that precede a given context next to each other in a permuted string, while another standard data structure the \emph{longest common prefix array (LCP)} then allows stretches of characters that precede contexts of a given length to be enumerated in a single pass through the two data structures. This enables the majority of scores to be smoothed to an average value, greatly improving compression.

We derive a formula to quantify the information lost during this smoothing process and justify our compression scheme empirically by showing that results using the compressed scores closely match the original data when our scheme is applied to whole-genome resequencing data. We also show that we can use the BWT alone to compress quality scores in a way that is almost as effective as BWT/LCP compression, thus avoiding the overhead of computing the LCP array. Moreover, we demonstrate that our methods can be used in tandem with other approaches to boost the compression obtained.

\section{Methods}
\label{sec:methods}
\subsection{Definitions} %

Given a string $s$ of $k$ symbols drawn from some finite ordered alphabet $\Sigma$, we mark the end of $s$ by appending an additional \emph{end marker} $\$$ such that $\$<c$ for any symbol $c$ in $\Sigma$. Starting at each position of $s$ and reading rightwards, we obtain $k+1$ distinct \emph{suffixes}. We say that each suffix is \emph{associated} with the character that precedes it in $s$ (one such suffix comprises the entirety of $s$, for which we ``wrap around'' and associate it with $\$$). Ordering the suffixes of $s$ alphabetically then replacing them by their associated symbols defines a permutation $s \rightarrow \mathsf{BWT}(s)$ of the symbols of $s$ known as the \emph{Burrows-Wheeler transform} (henceforth BWT) of $s$ (\cite{bwt94}).
Perhaps the two most important of its many interesting properties are that the BWT is \emph{reversible}, in the sense that $s$ can be reconstructed from $\mathsf{BWT}(s)$ with no additional information (\cite{bookBWTAdjeroh:2008}) and the \emph{clustering effect} of the produced output, \emph{i.e.} BWT tends to group together characters that occur in similar contexts in the input text, making the output more compressible even by simple compressors (for instance see \cite{RestivoRosoneTCS2011}).

A way to generalize the BWT to a set $S=\{s_1,s_2,...,s_n\}$ of strings is simply to append distinct end markers $\$_i$ to each $s_i$ such that $\$_1<...<\$_n <c$ for any $c$ in $\Sigma$. For a single string, the permutation $s \rightarrow \mathsf{BWT}(s)$ provides a relation to the \emph{suffix array} of $s$, which is defined by applying the same permutation to the integers $0, \ldots, |s| - 1$ so as to arrange the starting positions of the suffixes of $s$ into lexicographical order. The BWT of a collection is related to its \emph{generalized suffix array} in an analogous way. Formally, the GSA is defined such that $\mathsf{GSA}(S)[j]$ gives the position of the $j$-th smallest suffix of the strings in $S$, which is encoded as a pair $(t,i)$ denoting that the suffix starts at position $t$ of $s_i$. In particular, if $\mathsf{GSA}(S)[j]=(t,i)$ then $\mathsf{BWT}(S)[j] = s_i[(t-1) \mathrm{mod} |s_i|]$.

Now we suppose the elements of $S$ are DNA sequences that are accompanied by strings $Q=\{q_1,q_2,...,q_n\}$ such that the symbol $q_i[j]$ encodes the quality score of $s_i[j]$ - the alphabet used by $Q$ and the means of encoding are not relevant at this point. We apply the same permutation $S \rightarrow \mathsf{BWT}(S)$ to $Q$ (which we emphasize is not the same as computing the BWT of $Q$ itself) to obtain a string $QV$ such that $QV[i]$ encodes the quality score associated with the symbol $\mathsf{BWT}(S)[i]$.

The \emph{longest common prefix array} (denoted by LCP) of a collection $S$ of strings stores the length of the longest common prefixes between two consecutive suffixes of $S$ in the lexicographic order. For every $j=1, \ldots, n-1$, if $\mathsf{GSA}(S)[j-1]=(p_1,p_2)$ and $\mathsf{GSA}(S)[j]=(q_1,q_2)$, $\mathsf{LCP}(S)[j]$ is the length of the longest common prefix of suffixes starting at positions $p_1$ and $q_1$ of the words $s_{p_2}$ and $s_{q_2}$, respectively. We set $\mathsf{LCP}(S)[0]=0$.
Note that we do not need to compute explicitly the generalized suffix array in order to obtain the BWT and LCP. Methods suitable for computing $\mathsf{BWT}(S)$ and $\mathsf{LCP}(S)$ where $S$ is a large collection of DNA sequences and without the use of the GSA of $S$ were given in \cite{BauerCoxRosoneCPM11,BauerCoxRosoneTCS2012,BauerCoxRosoneSciortino2012} and it is straightforward to adapt them to compute $QV$ at the same time.\\

\subsection{Smoothing strategy}\label{subsec:smoothing}

A crucial consequence of the definition of the BWT is that the symbols associated with suffixes that begin with some string $w$ will form a contiguous substring in $\mathsf{BWT}(S)$, we call this the \emph{$w$-interval}. If all symbols in the $w$-interval take the same value $c$, then every occurrence of $w$ in $S$ is preceded by $c$: seeing $w$ in a read \emph{predicts} that $c$ will come before it.

For a fixed length $k$, a linear scan through the LCP array allows us to identify \emph{LCP-intervals}, which are maximal intervals $[i,j]$ that satisfy  $\LCP[r] \geq k$ for $i \leq r \leq j$ and whose associated suffixes therefore share at least the first $k$ bases. We set thresholds on the minimum length of the predicting context $k$ and the minimum number of times $j-i+1$ it must occur in $S$. If these are both exceeded and the symbols in $\mathsf{BWT}(S)[i,j]$ are all the same, then we smooth the corresponding quality scores in $QV(S)[i,j]$.

Since each score in $QV(S)[i,j]$ implies an error probability for its associated base, one way to do the smoothing would be to take the mean of these error rates across $QV(S)[i,j]$ and convert this to a score with which we replace all scores in $QV(S)[i,j]$ (which we note is not the same as taking the mean of the scores). However, better compression is obtained by replacing the smoothed scores with the score implied by the mean error rate of \emph{all} bases whose scores are smoothed. Moreover, we empirically observed that almost as good results are achieved just by smoothing all quality scores in $QV$ that are associated with runs of a given symbol in $\mathsf{BWT}(S)$ whose lengths exceed a threshold, which has the practical benefit of not needing the LCP array.

\subsection{Measuring the information loss due to smoothing}\label{subsec:entropy}

The probability distribution of a randomly chosen symbol from a string $s$ is $\textsf{p}(s)=(n_1/|s|,\ldots,n_{|\Sigma|}/|s|)$, where $n_1,\ldots,n_{|\Sigma|}$ count the occurrences of the symbols of $\Sigma$ in $s$. Applying the Shannon entropy transformation $H:(p_1,\ldots,p_n)\rightarrow - \sum_i p_i \log(p_i)$  to this distribution  (\cite{Shannon379_423}) yields
\begin{equation}\label{eq-H-0}
H_0(s)= H(\textsf{p}(s)) = - \sum_{i \in \Sigma} \frac{n_i}{|s|}\log \frac{n_i}{|s|},
\end{equation}
a quantity known as the \emph{zero-order empirical entropy} of $s$ (we assume $0 \log 0=0$ and adopt the convention that all logarithms are taken to the base 2).

Let $b(w,s)$ be the string formed by concatenating the symbols that immediately precede the occurrences of some substring $w$ of $s$. For a positive integer $k$ we define the \emph{$k$-th order empirical entropy} of $s$ as
\begin{align}\label{eq-H-k}
H_k(s) &=\frac{1}{|s|} \sum_{w \in \Sigma^k } |b(w,s)| H_0(b(w,s)) \nonumber \\
&=\frac{1}{|s|} \sum_{w \in \Sigma^k } |b(w,s)| H(\textsf{p}(b(w,s))).
\end{align}
This can be thought of as the mean entropy across all symbols of $s$ when a context of $k$ bases is taken into consideration. The computations of $H_0(S)$ and $H_k(S)$ for a collection $S$ are identical to the single-string case.
We note here the connection with $\mathsf{BWT}(S)$ and $\mathsf{LCP}(S)$: the strings $b(w,S)$ form disjoint substrings in $\mathsf{BWT}(S)$ whose order in the BWT matches the lexicographic order of their associated $k$-mers $w$ (see also \cite{Manzini2001}). 
The coordinates of the strings $b(w,S)$ in $S$ are precisely the LCP-intervals of length $k$, which we have observed can be computed by a single pass through $\mathsf{LCP}(S)$.

We can view $H_0(b(w,S))$ as the Shannon entropy of the distribution $\textsf{p}(b(w,S))$ obtained by assuming each symbol of $b(w,S)$ is exactly known (so that \emph{e.g.} a `\texttt{G}' corresponds to a distribution $p(\texttt{A},\texttt{C},\texttt{G},\texttt{T})=(0,0,1,0)$) and then computing the mean of these distributions across all symbols of $b(w,S)$. With this in mind, we can generalize $H_0(b(w,S))$ to imprecisely known symbols by replacing these exact symbol-level distributions with the ones that are implied by the quality values that are associated with the elements of $b(w,S)$. For example, a Q20  `\texttt{G}' receives a distribution $p(\texttt{A},\texttt{C},\texttt{G},\texttt{T})= (0.01/3,0.01/3,0.99,0.01/3)$ - we assume the three error bases are equiprobable. Taking the mean of these gives a new ``quality-aware'' distribution  $\textsf{q}(b(w,S),Q)$ for the symbol expected to precede $w$, from which we can compute a modified entropy  $H_0 (b(w,S), Q)=H(\textsf{q}(b(w,S),Q)$ via the Shannon formula. This in turn implies a generalization of the $k$-th order empirical entropy:%

\begin{align}\label{eq-HkSQ}
H_k(S,Q) &=\frac{1}{|S|} \sum_{w \in \Sigma^k } |b(w,S)| H_0(b(w,S),Q) \nonumber \\
&=\frac{1}{|S|} \sum_{w \in \Sigma^k } |b(w,S)| H(\textsf{q}(b(w,s),Q)).
\end{align}

We give two ways to describe the effect of smoothing $Q$ to obtain a new set of scores $Q'$. First, the improvement in compression is measured by comparing the size of the files produced when standard compression tools are applied to the BWT-permuted quality scores $QV$ and $QV'$. Second, the information loss is quantified by the \emph{relative entropy} (or \emph{Kullback-Liebler divergence}), which measures the information loss when a distribution $\textsf{p}'=(p'_1,\dots,p'_n)$ is used to approximate a distribution $\textsf{p}=(p_1,\ldots,p_n)$ and is defined by
\begin{equation}\label{KL}
\textrm{RE}(\textsf{p}||\textsf{p}')= \sum_i p_i \log{\frac{p_i}{p'_i}}.
\end{equation}
In much the same way as $H_k(S)$ is defined from $H_0(S)$, we may define the \emph{$k$-th order empirical relative entropy} $\textrm{RE}(Q||Q')$ as (see also \cite{EpifanioGabrieleGiancarloSciortino2011})
\begin{equation}\label{KL}
\frac{1}{|S|} \sum_{w \in \Sigma^k } |b(w,S)| \textrm{RE}(\textsf{q}(b(w,S),Q)||\textsf{q}(b(w,S), Q'))
\end{equation}
 Similar to $H_k(S)$, a single pass through $\mathsf{LCP}(S)$ allows us to compute $\textrm{RE}(Q||Q')$ by enumerating the $w$-intervals associated with each $k$-mer: this time, we need the $w$-intervals in $\mathsf{BWT}(S)$ plus the corresponding scores in $QV$ and $QV'$ to compute the terms $ \textrm{RE}(\textsf{q}(b(w,S),Q)||\textsf{q}(b(w,S), Q'))$.

Given two smoothings $Q \rightarrow Q'$ and $Q \rightarrow Q''$, the smaller of $\textrm{RE}(Q||Q')$ and $\textrm{RE}(Q||Q'')$ suggests the smallest loss in information content.
It can be verified that $\textrm{RE}(Q||Q') \ge 0 $ and that $\textrm{RE}(Q||Q')$ is only zero when the distributions $\textsf{q}(b(w,S),Q)$ and $\textsf{q}(b(w,S), Q')$ are identical for all $k$-mers $w$ in $S$.

The smoothing scheme we gave in Section~\ref{subsec:smoothing} can be understood in this context. If all of $b(w,S)$ support the same base call then it must be true that $\textsf{q}(b(w,S),Q)$ assigns some probability $p$ to that call and probability $(1-p)/3$ to the remaining bases. Replacing all the quality scores in $b(w,S)$ with the score corresponding to $p$ will create a smoothed set of scores $Q'$ for which   $\textsf{q}(b(w,S),Q)=\textsf{q}(b(w,S), Q')$ and hence $\textrm{RE}(Q||Q')=0$. In practice, the integer nature of the Phred scoring scheme means the value used to overwrite the smoothed scores will typically be a slight approximation to the score that corresponds to $p$. However, we found that $p$ did not vary much between different intervals $b(w,S)$ and that small changes to its value had little effect on results. We therefore opted to improve compression by replacing all smoothed scores with a globally chosen replacement value.

\section{Results}\label{sec:results}

\subsection{Parameter sweep}\label{subsec:sweeping}

Using \emph{C.elegans} data allowed us to study the effect of compression on variant calls made using whole-genome shotgun sequences from a diploid genome, while permitting more extensive parameter sweeps than would be tractable for human data. We chose a dataset SRR065390\_1 comprising 33,808,546 reads of length 100 ($33.6\times$ coverage of the genome) that has been previously studied by the Sequence Squeeze entrants.
For our computational pipeline we chose bwa\footnote{bwa version 0.6.1 with parameters -t 12 -q 15 (\cite{bwa2009})} followed by GATK\footnote{GATK version 1.6 (\cite{GATK2011})}.
Only the variants marked as ``PASS" by GATK (using UnifiedGenotyper and Variant Quality Score Recalibration) are considered.

We treat the results of this pipeline on uncompressed data as ``ground truth" (\emph{i.e.} we do not try to distinguish false positive or false negative calls in the uncompressed data, as we would on a simulated dataset) since we wish results derived from compressed data to reflect the original data as closely as possible.
We chose to measure the proportion of reads that were differentially mapped and the proportion of variant calls that were different.
We calculated sensitivity and specificity values and combined the two into an F-statistic.

\begin {figure}
{\scriptsize    %
$$          %
\includegraphics[bb = 60 50 400 274, width=60mm]{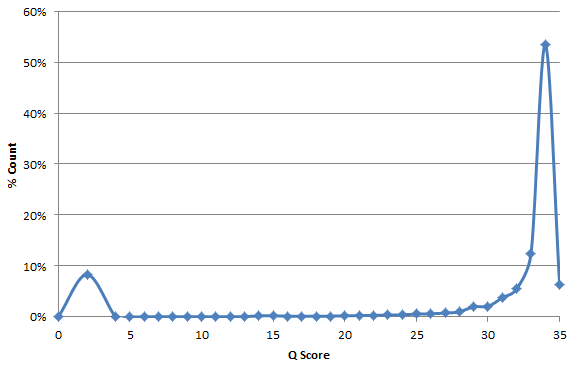}
$$
}
	\caption{Histogram of \emph{C. Elegans} SRR065390\_1 Q scores}
	\label{fig:HistogramQcelegans}
\end{figure}

Qualities are distributed as shown in Figure~\ref{fig:HistogramQcelegans}.
We notice an overrepresentation of Q2 scores (8.3\%) due to end-of-reads trimming.
A negligible number of `N' bases (0.07\%) are also associated to Q0 scores.
We verified that better results were obtained when these Q0 and Q2 scores were conserved, and all the results presented below keep these two Q scores unchanged.
To gain an idea of ``worst case scenario" behaviour, we first computed the mean estimate error rate by converting each quality score (ignoring Q0 and Q2) to an error probability, taking the mean of these values and converting back to Phred score  (which we note is not the same as taking the mean of the quality scores).
We obtained a mean Q score of 28.36 (as a comparison, the mean of the Q scores is 32.74).
We then replaced every Q score by this mean value and reran our pipeline.
This caused 15\% of variant calls to change (2.3\% missing calls and 12.7\% new variants calls), also characterised by the worst-case F-score of 92.8\%.

We then compressed the quality scores using various values for the two parameters of our quality smoothing algorithm: LCP cut threshold and minimum stretch length.
All stretches (in BWT-space) of same BWT letter longer than (or equal to) the minimum stretch length, and for which the LCP value stays above (or equal to) the LCP cut threshold gets ``smoothed", i.e. its associated qualities get reset to a constant quality score value (except for Q0 and Q2, which stay at their current values, but are still allowed to contribute to stretches).

In theory, as each set of parameters leads to different numbers of quality values being smoothed, the average Q score that should be used as a replacement shouldn't be constant across runs: in order to strictly follow the Q score definition associated with error rates, we recalculated for each run the Q score that would lead to the same mean error rate, and used it as a replacement.
However, we noticed that it only varied between 28.36 and 30.46 and that its effect on variant calling was negligible.
For this reason, the results presented below are using a fixed replacement Q score of 29 for all runs.

Figure~\ref{fig:SmoothingResults} summarizes the results obtained for a sweep of parameters.
In this table, the results of the ``LCP-free" smoothing appears when ``LCP cut threshold=0".
For each combination of parameters, we compressed the BWT-ordered smoothed qualities using \bzip2\ and \7zip's PPMd algorithm\footnote{7-zip  with parameters -m0=PPMd ; http://www.7-zip.org}.
\7zip compression was always better and is the one reported in the table, in bits per quality value, in BWT-space (column 3) and in read-space (column 4).
We also calculated F-scores (column 5) based on the number of false positive and false negative variant calls.

\begin{figure}[t]
\begin{threeparttable}[b]
    \begin{tabular}{|l|l|l|l|l|l|}
        \hline
        LCP cut  & Min  & \%Q3+ & BWT- & Read- & Variant \\
         threshold & stretch  & cut & space & space & calling \\
          ~ & length & ~ & compress. & compress. & F-score \\
          ~ & ~  & ~ & (bits/Q)\footnotemark[6] & (bits/Q)\footnotemark[7] & ~ \\ \hline
        uncut         & ~              & 0\%     & 2.51  & 1.67   & 100\%          \\
        10            & 10             & 76.7\%     & 1.28 & 1.00    & 99.1\%     \\
        5             & 10             & 76.8\%     & 1.28  &  0.99 & 99.1\%       \\
        0 \footnotemark[9]            & 10             & 76.8\%     & 1.28  & 0.99   & 98.8\%      \\
        5             & 5              & 85.9\%     & 1.06  & 0.68   & 97.8\%        \\
        1             & 5              & 86.1\%     & 1.06  & 0.68   & 97.9\%        \\
        0 \footnotemark[9]            & 5              & 86.1\%     & 1.06  & 0.68   & 97.7\%        \\
        5             & 1              & 96.9\%     & 0.50  & 0.20   & 92.3\%        \\
        1             & 1              & 99.3\%     & 0.39  & 0.11   & 92.9\%        \\
        0 \footnotemark[8]\footnotemark[9]            & 1              & 100\%     & 0.37   & 0.06  & 92.8\%         \\
        \hline
    \end{tabular}
  \begin{tablenotes}
    \item[6]{compression in BWT-space in bits per quality value}
    \item[7]{compression in read-space (i.e. same as FastQ file) in bits per quality value}
    \item[8]{all cut except Q0 and Q2, which are kept intact in all cases}
    \item[9]{LCP cut threshold = 0 doesn't use LCP}
  \end{tablenotes}
 \end{threeparttable}
 \caption{Statistics after quality smoothing}
 \label{fig:SmoothingResults}
\end{figure}

We notice that minimum stretch length is the parameter having the most influence on both the compression rate and the fidelity of variant calling.
A minimum stretch length of 10, meaning that 10 consecutive identical BWT letters must share the same following $k$-mer (whose length depends on the LCP cut threshold) in order to get smoothed, already leads to 76\% of the qualities getting replaced (excluding Q0 and Q2).
 In this situation, the quality string in BWT-space gets compressed at 1.28 bits per quality value, and in read-space (i.e. same as the original FASTQ) at 1 bit per base.
The effect on variant calling is minimal, as the F-score of 99\% corresponds to 30 variants getting missed (i.e. false negatives) from the original 3208 called and 40 variants being created (i.e. false positives).
However, in all the cases, the false negative and false positive calls are present in the opposite variants file but didn't pass filter due to a quality slightly below the required threshold.

The LCP cut threshold has a very marginal effect on both the compression rate and the variant calling.
In fact, it has an unpredictable effect on variant calling, where a reduction of the threshold sometimes lead to slightly better and sometimes to slightly worse results.
An interesting observation is that reducing the LCP cut threshold to zero, \emph{i.e.} when the LCP value is not used at all, doesn't affect the results as negatively as we expected: in this situation, any stretch of consecutive identical BWT letters gets smoothed even if the letters don't share a common following $k$-mer (as the minimal length of this $k$-mer is zero).
This is explained by the relatively low number of occurrences of such stretches: Figure~\ref{fig:SmoothingResults} column 3 reveals that only 0.2\% of the bases get affected by a change of LCP cut threshold from 5 to 0 in the most interesting case where min stretch length is 5.

Another interesting observation is the better compression obtained in read-space than in BWT-space.
This is explained by the tendency for bases to stay constant from one cycle to the next (a property observable in read-space), but don't have any reason to stay constant across reads sharing the same suffix.
This makes compression in BWT-space more difficult.

\begin{figure}[t]
\begin{threeparttable}[b]
    \begin{tabular}{|l|l|l|l|}
        \hline
         Smoothing  & BWT-space & Read-space & Variant \\
         strategy & compression & compression & calling \\
          ~   & (bits/Q) & (bits/Q) & F-score \\ \hline
        Original cut5 stretch5      & 1.06  & 0.68   & 97.8\%        \\
        Random same dist.      & 1.05  & 1.09   & 94.8\%        \\
        \hline
    \end{tabular}
 \end{threeparttable}
 \caption{Comparison with random smoothing of same stretch lengths distribution}
 \label{fig:ComparisonWithRandomSmoothing}
\end{figure}

We also wished to verify that randomly smoothing qualities had a more detrimental effect on variant calling.
Based on the distribution of smoothed stretch lengths obtained with LCP cut threshold = 5 and minimal stretch length = 5, we smoothed random stretches of the original dataset in such a way as to achieve the same final distribution.
After running our computational pipeline on this randomly smoothed dataset, we obtained the statistics shown in Figure~\ref{fig:ComparisonWithRandomSmoothing}: same compressibility in BWT-space, but much worse in read-space, and worse variant calling F-score. The lower compressibility in read space is explained by the fact that the qualities being replaced are now distributed less consecutively than before in read space: our smoothing strategy, even though applied in BWT space, is occurring at specific places that are permuted into consecutive positions in read space. Instead, the smoothing of random BWT stretches doesn't get permutated into consecutive positions in read space and leads to a worse compression rate. It can be noted that some compression still happens because 85.9\% of the qualities, which had various values before smoothing, are reduced to a single value and therefore become more compressible.

\begin{figure}
{\scriptsize
$$
	\includegraphics[bb = 140 140 620 520, width=60mm]{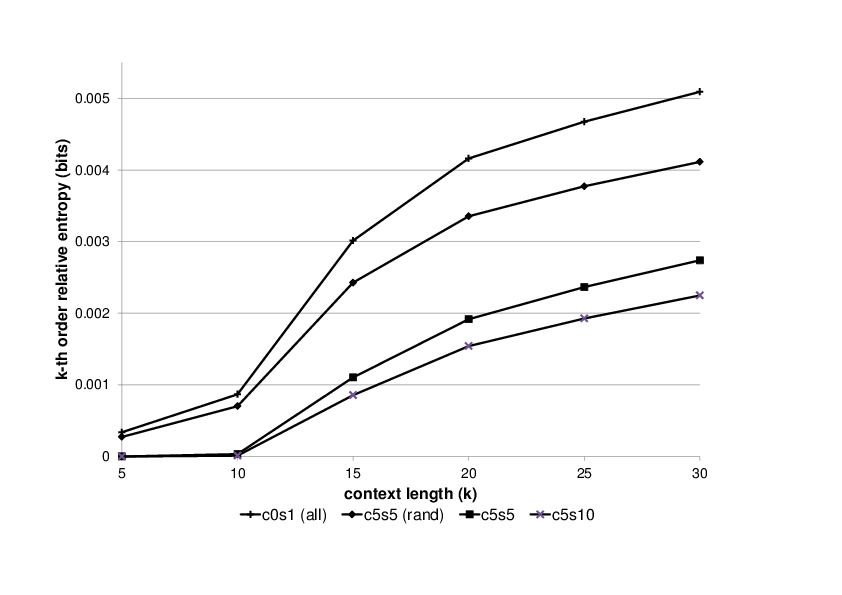}
$$
}
	\caption{$k$-th order empirical relative entropy between the original and compressed quality scores of SRR065390\_1, for various quality-score compression schemes}
	\label{fig:Entropy}
\end{figure}	

Finally, Figure~\ref{fig:Entropy} depicts the $k$-th order relative entropy between the full-resolution quality scores and a subset of the compression schemes mentioned in Figures~\ref{fig:SmoothingResults} and~\ref{fig:ComparisonWithRandomSmoothing}. We would expect the worst-case information loss to occur when all quality scores are replaced by a constant value and the `c0s1' curve behaves accordingly, exhibiting the highest relative entropy. The `c5s5' and `c5s10' curves correspond to the `stretch length 5' and `stretch length 10' entries in Figures~\ref{fig:SmoothingResults} for an LCP cut threshold of 5. As we would expect, the more aggressive of the two schemes `c5s5' has a higher relative entropy with respect to the original scores, suggesting a greater loss of information. The curve  for `c5s5' contrasts with the `c5s5 (rand)' curve for the randomly-smoothed dataset compared against it in Figure ~\ref{fig:ComparisonWithRandomSmoothing}. While Figure ~\ref{fig:ComparisonWithRandomSmoothing} shows nearly identical compression for the two datasets, Figure~\ref{fig:Entropy} reveals that randomly smoothing the quality scores has caused a much greater loss of information.

\subsection{Details and requantization}\label{subsec:comparison}

The F-scores presented in the previous section, although good at summarising the overall impact of compression on variant calling, are abstracting away some important details.
In this section, we use QQ plots to show the correlation between variant call qualities (QUAL field reported by GATK) with and without compression.
These QQ plots are also highlighting some interesting differences between our BWT-based compression strategy and the requantization strategy (reduction from 40 to 8 quality scores) at similar compression and F-score rates.

\begin{figure}
{\scriptsize
$$
	\includegraphics[bb = 70 70 430 450,width=60mm]{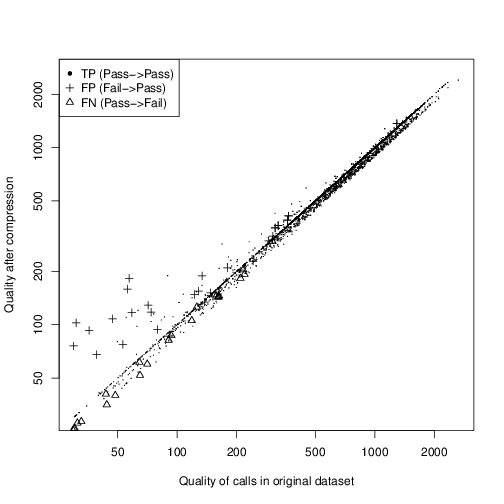}
$$
}
	\caption{Correlation of variant call qualities before/after c5s10 compression}
	\label{fig:QQplot_c5s10}
\end{figure}

Considering our smoothing strategy with parameters \{cut threshold = 5, min stretch length = 10\}, which was leading to a compression rate of 1.28 bits/Q in BWT space and 0.99 bit/Q in read space, and was associated to a F-score of 99.1\%, Figure ~\ref{fig:QQplot_c5s10} shows for each variant called its quality before (X axis) and after compression (Y axis).
We distinguish three classes of variants: True Positives (TP) are those called as "passing filter" (as defined by GATK) before and after compression; False Positives (FP) are those passing filter after compression but not before; False Negatives (FN) are those passing filter before compression but not after.
FP and FN calls, although passing filter in only one of the two GATK runs, always happen to be present in both call sets, allowing us to plot their quality values.

Figure ~\ref{fig:QQplot_c5s10} shows 36 FP, 25 FN and 3183 TP calls. We notice that the majority of stray points are FP calls whose quality has been enhanced by our algorithm.
In fact, all the qualities considered in this QQ plot, before and after compression, are above the filter threshold of quality 30 (the X and Y axis start at Q=30).
The reason for calls not passing filter in the original dataset is most often due to the LowQD filter, but a direct link to the aligner's mapping quality of reads or number thereof has not been established.

\begin{figure}
{\scriptsize
$$
	\includegraphics[bb = 70 70 430 450,width=60mm]{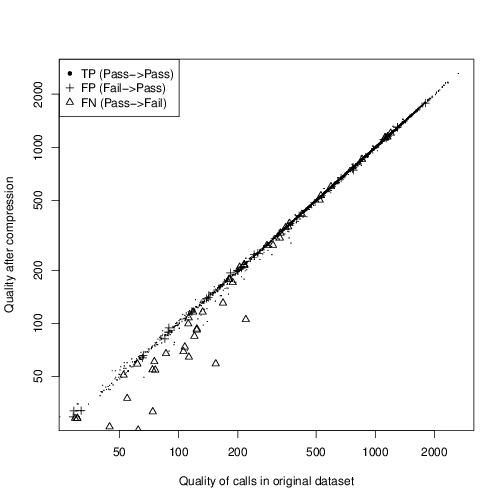}
$$
}
	\caption{Correlation of variant call qualities before/after requantization}
	\label{fig:QQplot_8bin}
\end{figure}

For comparison with another compression strategy, Figure ~\ref{fig:QQplot_8bin} shows the QQ plot obtained after requantization of the original dataset to 8 Qscore bins. This led to the same compression rate as the c5s10 strategy (which was in fact chosen for this reason): 1.29 bits/Q in BWT space.
The F-score of quantization, 89.9\%, was also very similar to c5s10's.
However, quantization achieves this F-score with 27 FP, 51 FN and 3157 TP calls, which is much more biased towards FN, whereas c5s10 (and all our other observed results from this smoothing strategy) was more pronounced towards FP.

Due to the dataset used in this paper, we haven't reached any conclusion regarding the quality of those FP and FN calls: some FP calls may introduce real false positive calls, whereas others may reveal some real variants that hadn't been called in the original dataset.
Similarly, quantization's FN calls may be correct pruning of previously incorrect calls as well as real false negatives.
We intend to run the same analysis on a simulated dataset where we will have prior knowledge of the correctness of calls.

\begin{figure}
{\scriptsize
$$
	\includegraphics[bb = 70 70 430 450,width=60mm]{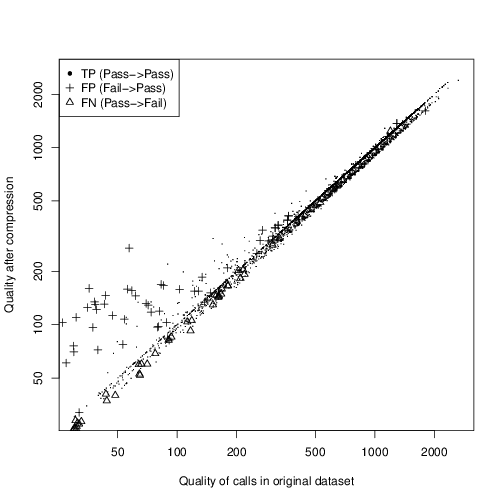}
$$
}
	\caption{Correlation of variant call qualities before/after c0s5 compression}
	\label{fig:QQplot_c0s5}
\end{figure}

Finally, Figure ~\ref{fig:QQplot_c0s5} shows the distribution of 108 FP, 44 FN and 3164 TP obtained with c0s5 smoothing.
Relatively to c5s10, the number of FP increases faster than the number of FN and this trend continues when we push the smoothing parameters towards lower min stretch size and higher compression rates.
Also, confirming an expected behaviour, the FP calls present after c5s10 smoothing are still present as FP calls after the more compressed c0s5 smoothing.

\section{Discussion}\label{sec:discussion}
This paper aims to introduce the general idea of smoothing quality scores based on the BWT and LCP of their associated reads. Section~\ref{subsec:smoothing} describes perhaps the simplest and most conservative approach to this, but the theoretical framework presented in Section~\ref{subsec:entropy} allows comparison of future more sophisticated quality smoothing strategies. Note that the effect of smoothing or otherwise adjusting the quality scores is to change the weightings of the different nucleotides in the distributions $X(w, QV)$. We can take a step further and consider adjusting low-probability bases in $X(w, QV)$ to zero. Our work can thus be extended to a new quality-based view on \emph{de novo error correction} that may provide an interesting alternative to existing approaches, many of which are based on the counting of $k$-mers (as surveyed by \cite{Yang2013}). Such a strategy could be thought of as a quality-aware extension of the HiTEC algorithm (\cite{Ilie2011}), while enjoying the considerable space advantage of being based on a (potentially compressed) BWT instead of a suffix array.

While it is an advantage that the relative entropy measures the information lost by quality smoothing in an application-neutral way, we also recognise that a single numerical quantity cannot fully model the effect of quality smoothing on the often complex multi-step analysis pipelines that are applied to sequence data. We investigated this by measuring the effect of smoothed quality scores on the results of widely-accepted tools for a well-understood application.

Transforming the smoothed scores back into their original reads added a significant boost to the compression already achieved by \7zip from exploiting similarities between the quality scores of individual read. It is equally simple to combine our approach with the application to the unsmoothed scores of one of the lossy requantization schemes studied by \cite{Wan2012}.

However, using our method in this way involves building the BWT (and possibly LCP) of a set of reads and then applying the inverse BWT permutation to the quality scores to obtain a smoothed quality string for each read. The overhead of these tasks may limit its practicality for downstream applications that operate on reads and does not utilise the potential of the BWT.
As well as allowing excellent lossless compression, storing sequences in BWT form also facilitates rapid analysis: the sga (\cite{Simpson2011}) and Fermi (\cite{Li2012}) assemblers both operate directly on BWT-based compressed indexes of sets of reads and we ourselves have shown that similar data structures can be the basis of both RNA-Seq (\cite{Cox2012b}) and metagenomic (\cite{Ander2013}) analyses.
While the compression achieved on BWT-space reads is less than in read-space (although  the difference may be less clear-cut on a dataset where the Q2-masking of read ends is less prevalent or has been switched off), we therefore envisage that a key application of our work is to allow quality scores to be used in a BWT-space context while being stored in as compact a manner as the reads themselves.

\paragraph{Funding\textcolon} L.J. and A.J.C. are employees of Illumina Inc., a public company that develops and markets systems for genetic analysis, and receive shares as part of their compensation.

\bibliographystyle{natbib}
\bibliography{BWT}

\end{document}